\documentclass[prl,showpacs, twocolumn,groupedaddress]{revtex4}

\usepackage{graphicx}

\begin{document}

\title
{ Coulomb interactions of massless Dirac fermions in graphene;
 pair-distribution functions and exchange-driven spin-polarized phases.
}

\author
{ M.W.C. Dharma-wardana
}
\email[Email address:\ ]{chandre.dharma-wardana@nrc.ca}
\affiliation{
National Research Council of Canada, Ottawa, Canada. K1A 0R6\\
}

\date{22 April 2006}
\date{\today}
\begin{abstract}
The quasi-2D electrons in graphene
behave as massless fermions obeying a Dirac-Weyl equation in the low-energy
regime near the two Fermi points.
 The stability of spin-polarized phases (SPP)
in graphene is considered. The exchange energy is evaluated
from the analytic pair-distribution functions, and the correlation
energies are estimated via a closely similar
four-component 2D electron fluid which has been investigated
previously.
SPPs appear for sufficiently high doping,
 when the exchange energy
alone is considered. 
However, the inclusion of correlations
is found to
{\it suppress } the spin-phase transition
in ideal graphene. 
\end{abstract}
%
\maketitle
%
{\it Introduction--}
The carbon atoms in graphene form a quasi-two-dimensional (Q2D) honeycomb
 lattice and contribute one electron per carbon to form an unusual 2D electron
system (2DES) with a massless Dirac-fermion dispersion near
 the Fermi points\cite{wallace,ando}.
 Graphene and related materials (e.g, nanotubes, fullerenes)
 have become a mine of  novel technologies and a new paradigm for studying
various aspects of physics\cite{geim}, including cosmological models on
honeycomb branes,
 superconductivity on bi-partite lattices\cite{nagamatsu} 
 and nanotubes\cite{cdw}, Hubbard models\cite{tosatti},
 spin-phase
transitions\cite{pgcn},
 and other
aspects of strongly correlated electrons\cite{khvesh}.

The hexagonal Brillouin zone has two inequivalent
points {\bf K}=$(1/3,1/\surd{3})$ and {\bf K$^\prime$}=$(-1/3,1/\surd{3})$,
in units of $2\pi/a_0$, where $a_0$ is the lattice constant. 
The simplest tight-binding model with nearest-neighbour hopping $t$
is sufficient to describe the low-energy regime where the valence
and conduction bands ($\pi$ and $\pi^*$) have linear dispersion
 near the {\bf K, K$^\prime$} points, with
zero bandgap. The graphene 2D electron system (G2DES)  
is nominally ``half-filled'', with the $\pi^*$ band unoccupied,
 and has spin and
valley degeneracies, with a Berry phase associated with the
 valley index\cite{ando}.

The vanishing of the density of states and the effective mass of the
2D electrons near the Fermi points suggest that the 
Coulomb interactions of the massless fermions remain
strong, unlike in the usual Fermi-liquid picture.
This also means that perturbation approaches have
to be treated with great caution. The Coulomb
interactions may induce a gap between the $\pi$ and $\pi^*$ bands,
or a lifting of the sublattice (valley) degeneracies, or stabilize
spin-polarized phases (SPP) in preference to the unpolarized state.
Such SPPs in GaAs/AlAs  2DES, predicted to appear at low coupling ($\sim 2-4$)
when perturbation methods are used, get pushed to
 high coupling if non-perturbative approaches 
were used($\sim ~26$), as discussed in Refs.~\cite{attac,prl3,2valley}.
 The two-valley
2D electron system (2v-2DES)  does {\it not} show a SPP when full non-local
non-perturbative calculations are used,
 presumably because these methods properly account for
 the direct interactions
 which are three times as  many as the exchange interactions\cite{2valley}.
The exchange and correlation energy $E_{xc}$ in the 2v-2DES of the
Si/SiO$_2$ system was calculated from the distribution functions
in Ref.~\cite{2valley},
using the classical-map hyper-netted-chain (CHNC)
technique, accurately recovering the Quantum Monte-Carlo (QMC) results
even in the strong coupling regimes\cite{conti}. CHNC provides the
pair-distribution functions (PDFs) $g_{ij}(r)$ as a function of the coupling
strength.
Then $E_{xc}$ is evaluated
via a coupling constant integration.
The method has been successfully applied to a variety of problems
including the 2DES\cite{prl2},
the 2v-2DES in Si-MOSFET devices\cite{2valley},
 and the thick quasi-2DES in HIGFET structures\cite{q2des}.
  However, a direct calculation of
exchange and correlation in graphene involves and $8\times8$ matrix of
two-component PDFs due to the spin and valley indices as well as the
presence of $\pi,\pi^*$ bands. Hence in this study we consider the
exchange energy $E_x$ via an analytic evaluation of the non-interacting
PDFs, and indirectly estimate the correlation energy $E_c$ using 
results for the spin-polarized four-component 
2v-2DES, exploiting the analogies between the two systems.
 
 The non-interacting PDFs of G2DES, viz.,
$\mathcal G^0_{ij}(r)$
 involve two components, the first being
a Bessel function as in the ordinary 2DES, while a second,
 associated with the cosine of the angle of 
e-e scattering, involves Bessel and Struve functions, as we show below.
We find that there are stable SPPs in an exchange-only approach.
However, including the correlation energy using the 2v-2DES data
stabilizes the {\it disorder-free}
G2DES in the unpolarized state. 
\begin {figure}
\includegraphics*[width=8.0cm, height=9.0cm]{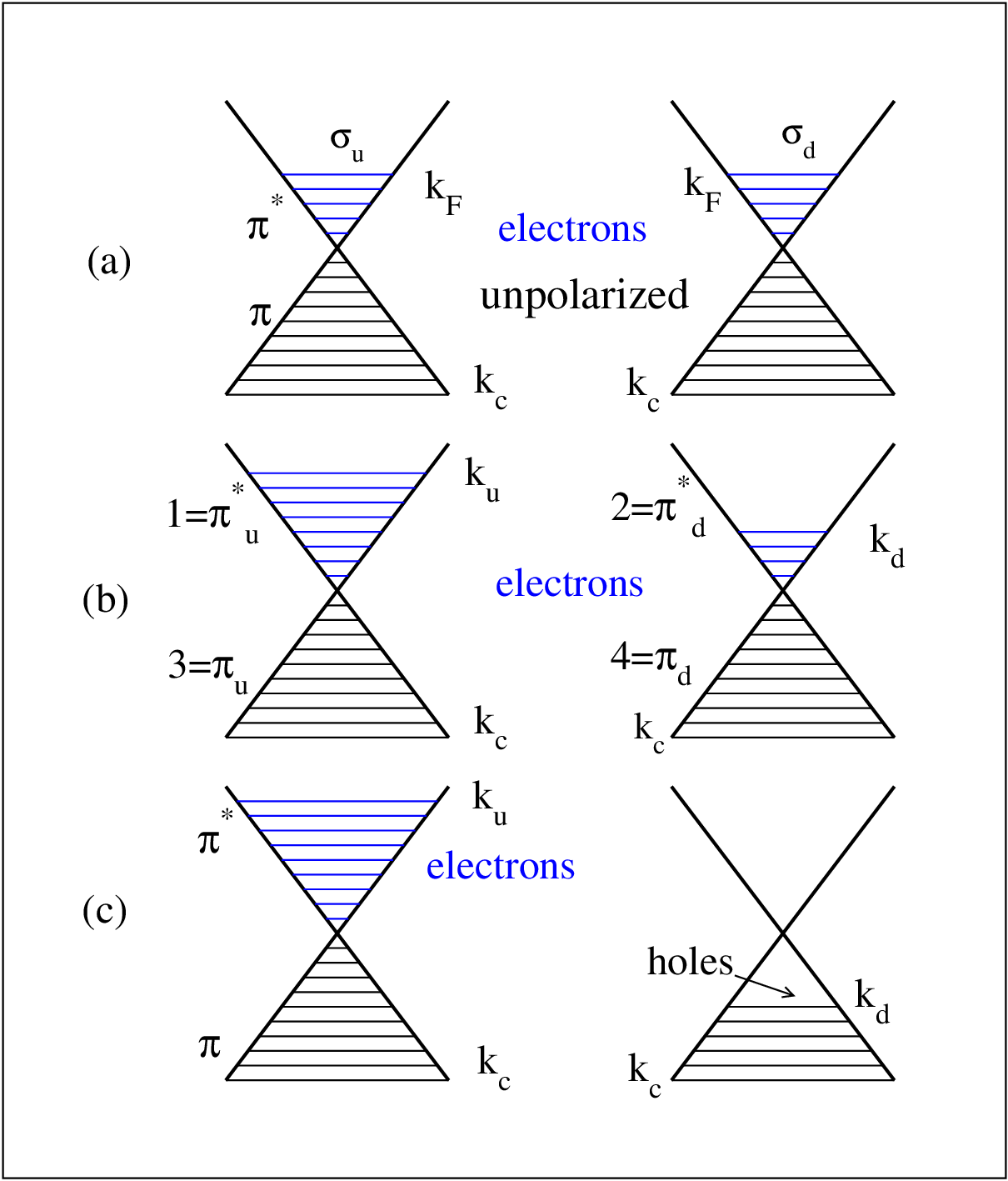}
\caption{
(Color online)Linear dispersion bands near a {\bf K} point where the
$\pi^*$ and $\pi$ bands cross. In (a) we show a doped unpolarized
system with equal occupation of the up-spin ($\sigma_u$) and
down-spin states. In (b) the polarized system has only electron
carriers. In (c) both electron and hole carries occur. This is
the only possibility if doping is zero ($k_F=0$ and $k_u=k_d$).
}
\label{figcones}
\end{figure}

{\it Model for the graphene 2DES --}The kinetic energy near
the {\bf K} points is given by a Dirac-Weyl Hamiltonian of the form: 
\begin{equation}
\label{hamil}
H_k=v_F(p_x\tau_z\sigma_x+p_y\sigma_y)
\end{equation}
Here $\tau_z=\pm 1$
defines the degenerate valleys, and $\sigma_x,\, \sigma_y$ denote
 the $x$ and $y$ Pauli
matrices that act in the space of the two atoms in each unit cell.
The $\pi,\pi^*$ bands of
spin and valley degenerate states (Fig.~\ref{figcones})
show a linear dispersion $E=\pm v_F\hbar k$.
 This form requires
a cutoff momentum $K_c$ such that the number of states in the
Brillouin zone is conserved. That is, if $A_0$ is the 
area per carbon, then $K_c^2=4\pi (1/A_0)$. The electron density
$N_c$ at half-filling is $1/A_0$, with $A_0=a_0^2\surd{3}/2$,
since one $\pi$ electron of arbitrary spin is
provided by each carbon atom. 
The Fermi velocity $v_F=ta_0\surd{3}/2$ is
thus the slope of the linear dispersion, with  $v_F\sim$ 5.5 eV\AA.
If the G2DES is embedded in a medium with dielectric
constant $\epsilon_0$, then we define
\begin{equation}
g^0=\frac{e^2/\epsilon_0}{\hbar v_F}=\frac{e^2}{\epsilon_0a_0}
/(t\surd{3}/2).
\end{equation}
This is the ratio of a typical Coulomb energy 
to the hopping energy and hence is usually taken as the Coulomb coupling
constant of the G2DES.  
This plays the same role as the $r_s$ parameter
in electron-gas theory of nonrelativistic finite-mass fermions.
The usual $r_s$ is not available for
G2DES since the effective mass $m^*$ is zero and there is no effective
Bohr radius. The coupling constant $g^0$ is {\it maximized} if $\epsilon_0$
is unity, and consistent with this case
 we assume $g^0=2.75$ for our G2DES studies, and do not treat it as tunable.

The 4-component envelope-eigenfunctions of the kinetic energy term are
made up of two-component functions $U=(b,e^{i\phi_k})$,
$U'=(e^{i\phi_k,b})$ and  $O=(0,0)$
where $\phi_k$ is the angle of the vector $\vec{k}$ in the 2-D plane.
Thus
\begin{eqnarray}
F^{\bf K}_{b,\vec{k}}(r)&=&(2A)^{1/2}(U,O)_T\chi_\sigma\\
F^{\bf K^\prime}_{b,\vec{k}}(r)&=&(2A)^{1/2}(O,U')_T\chi_\sigma
\end{eqnarray}
Here $b=\pm 1$ is a $\pi^*,\pi$ band index, $(\cdots)_T$ indicates the
transpose, and $\chi_\sigma$ is the spin function.
Then, using $v=1,2$ as a valley index, the Coulomb interaction
may be written in the form:
\begin{eqnarray}
\nonumber
H_I&=&\frac{1}{8A}\sum_{v_i,b_i,\sigma_i}\sum_{{\bf k,p,q}}V_q
\left[b_1b_4e^{i\{\phi^*({\bf k})-\phi({\bf k}+\bf{q})\}}+1\right]\\
\nonumber
 & &\times\left[b_2b_3e^{i\{\phi^*({\bf p})-\phi({\bf
 p}+\bf{q})\}}+1\right]\times\\
 & &a^+_{{\bf k},v_1,b_1,\sigma_1}a^+_{{\bf p}+{\bf q},v_2,b_2,\sigma_1}
a_{{\bf p},v_2,b_3,\sigma_2}a_{{\bf k}+{\bf q},v_1,b_4,\sigma_2}
\end{eqnarray}
Here $a^+,a$ are electron creation and annihilation operators and
$V_q=2\pi e^2/(\epsilon_0 q)$ is the 2D Coulomb interaction. The
phase factors introduce a novel $\cos(\theta)$ contribution where
$\theta$ is the scattering angle, not found in the usual jellium-2DES.
The resulting form of the exchange energy per Carbon is:
\begin{eqnarray}
E_{x}/E_u&=&-\frac{A_0g^0/k_c}{(2\pi)^2}\frac{1}{4}\sum_{b_1,b_2,\sigma}
\int_0^{2\pi} d\theta dk dp \\
\nonumber
 & &\times kp\frac{1+b_1b_2\cos(\theta)}{|{\bf k}-{\bf p}|}
n_{b_1,\sigma}(k)n_{b_2,\sigma}(p)
\end{eqnarray}
Here we have introduced the intrinsic coupling constant $g^0$ and the
energy unit $E_u=v_Fk_c$. Here $k_c=K_c/\surd{2}=\surd(4\pi n_c)$ is
based on the electron density per spin species, $n_c=N_c/2=1/(2A_0)$.
The above form of the exchange energy
can be reduced to an evaluation of a few elliptic integrals\cite{pgcn}.
The normal ``half-filled'' G2DES can be doped with electrons or
holes; but it is easy to show that symmetry enables us to limit to
one type of doping. However, given a system 
with an areal density of $N_\delta$
dopant electrons per valley, with $n_\delta=N_\delta/2$
per spin, the carriers in the spin-polarized
system could be electrons only,
 or both electrons and holes,
as shown in Fig.~\ref{figcones} for the $\pi^*$ and $\pi$ bands at
one {\bf K} point.
 The intrinsic system with $n_\delta=0$ can  be
an unpolarized state, as in Fig.~\ref{figcones}(a),
 or spin-polarized state with
electrons {\it and} holes, as in Fig.~\ref{figcones}(c).
 Such exchange-driven systems have
been studied by Peres et al.\cite{pgcn}, while the
correlation
effects have not been considered. Since the correlation
energy terms out-number the exchange terms by 3:1, it
is imperative to include the correlation energies.
The calculation of correlation energies is always
more difficult than exchange energies.
Peres et al. treated $g^0$ as
an externally tunable parameter for forcing SPPs.
Here we evaluate the exchange energy $E_x$ at the intrinsic $g^0$,
from the non-interacting
PDFs, and indirectly evaluate the correlation energy $E_c$ from
the four-component 2v-2DES
with the same coupling strength ($r_s=g^0$) and spin-polarization.  
\begin {figure}
\includegraphics*[width=7.0cm, height=8.0cm]{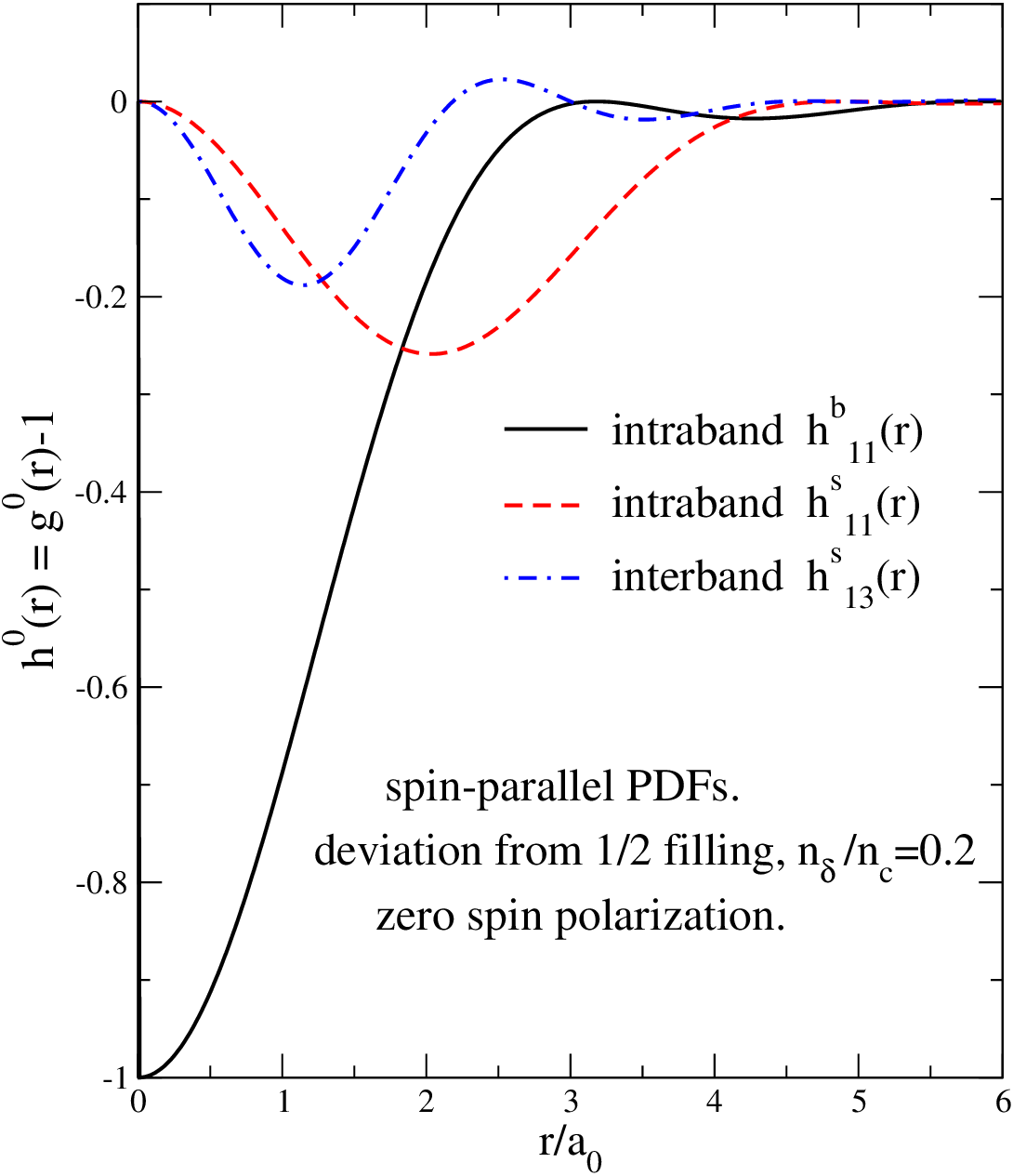}
\caption
{(Color online)The Bessel-like and Struve-like non-interacting,
 parallel-spin PCFs
$h^b(r)$ and $h^s(r)$ for the unpolarized doped system. The
bands are numbered as in Fig.~\ref{figcones}(b). The  anti-parallel
non-interacting PCFs are zero.
}
\label{figgr}
\end{figure}

{\it Pair-distribution functions of the G2DES--}
Although we are dealing with a four-component
system (2-valleys, 2-spin states), as seen from Fig.~\ref{figcones},
we need to consider the redistribution
of electrons and holes among the $\pi^*$ and $\pi$ bands when
comparing the energy of spin-polarized states with the
unpolarized
state. From Fig.~\ref{figcones} we see that the
e-e interactions at a given valley can be constructed from
(i) interactions with 
a $\pi^*\sigma_u$ band of up-spin electrons of density $n_u$,
 filled to $k_u$,
 (ii) a $\pi^*\sigma_d$
 set of electrons 
or a $\pi\sigma_d$ spin-down holes, of density $n_d$,
filled up to $k_d$ 
(iii)the $\pi\sigma_u$ band, with electron density $n_c$,
 filled to $k_c$ and 
(iv)the $\pi\sigma_d$ band, density $n_c$, filled to $k_c$
There will also be similar inter-valley terms.  
Each term in this 4$\times$4 matrix, denoted by $\mathcal G_{ij}(r)$
where $i$, $j=1\cdots4$,
 will have two components
associated with those in $U'$ and $U=(b,e^{i\phi_k})$.
Thus $\mathcal G_{ij}(r)=g^b_{ij}(r),g^s_{ij}(r)$, where the
superfixes ``$b,s$'' indicate that the noninteracting forms are
Bessel-function like, and Struve-function like, respectively. The
Struve form arises from the $\cos(\theta)$ terms in
the Coulomb interaction.
 The
numbering scheme of the matrix is shown in Fig.~\ref{figcones}(b).
Thus, defining the pair-correlation functions (PCFs)
$\mathcal H_{ij}(r)=\mathcal G_{ij}(r)-1$, or its components
$h_{ij}(r)=g_{ij}(r)-1$, we have:
\begin{eqnarray}
\nonumber
h^{0,b}_{ij}(r)&=&-(n_in_j)^{-1}\int_0^{k_i}\frac{d{\bf k_1}}{(2\pi)^2}
\int_0^{k_j}\frac{d{\bf k_2}}{(2\pi)^2} 
e^{i({\bf k}_1-{\bf k}_2)\cdot{\bf r}}\\ 
 &=&-\frac{2}{k_ir}J_1(k_ir)\frac{2}{k_jr}J_1(k_jr) \\
 \nonumber
h^{0,s}_{ij}(r)&=&-(n_in_j)^{-1}\int_0^{k_i}\frac{d{\bf k_1}}{(2\pi)^2}
\int_0^{k_j}\frac{d{\bf k_2}}{(2\pi)^2}\\
\nonumber
 & &\cos(\theta_1-\theta_2)
 e^{i({\bf k}_1-{\bf k}_2)\cdot{\bf r}}\\
\nonumber 
&=&-\frac{\pi}{k_ir}\frac{\pi}{k_jr}[J_0H_1-J_1H_0]_i
[J_0H_1-H_0J_1]_j 
\end{eqnarray}
Here $J_0,J_1$ are Bessel functions, while $H_0$ and $H_1$
are Struve $H$-functions. Also, in $[J_0H_1-J_1H_0]_i$ the
functions are evaluated at the argument $k_ir$. The wavevectors
$k_i=\surd{(4\pi n_i)}$ are for each component $i$, of density $n_i$.
We show typical noninteracting PCFs for a doped, unpolarized case 
as in, Fig.~\ref{figcones}(a), with the doping fraction $n_\delta/n_c=0.2$.
In CHNC, the exchange-hole is mapped exactly into
a classical Coulomb fluid using the Lado procedure\cite{prl2}. 
The figure shows that the exchange-hole is strongly reduced by the
presence of the $\cos(\theta)$ term which has been averaged into the
Struve-like PCFs $h^s(r)$. When the Coulomb interaction is included,
the $\cos(\theta)$ term has a similar mitigating effect and
 exchange-correlation in the G2DES is actually 
{\it considerably weaker} than in the
corresponding 2v-2DES which may be realized in a
Si MOSFET. The CHNC calculation for the
2v-2DES for the conditions stipulated in Fig.~\ref{figgr} show that
the correlation energy is only about a third of the exchange energy
at $r_s=g^0$.
This and other model calculations allow us to conclude that a good
approximation is to ignore the $E_c$ contribution arising from the
Struve-like distribution functions and estimate the $E_c$ only from
the Bessel-like functions which are the only forms appearing in the
2v-2DES. This justifies our use of the 2v-2DES for the correlation
energy, while the $E_x$ is exactly evaluated. 

\begin {figure}
\includegraphics*[width=8.0cm, height=10.0cm]{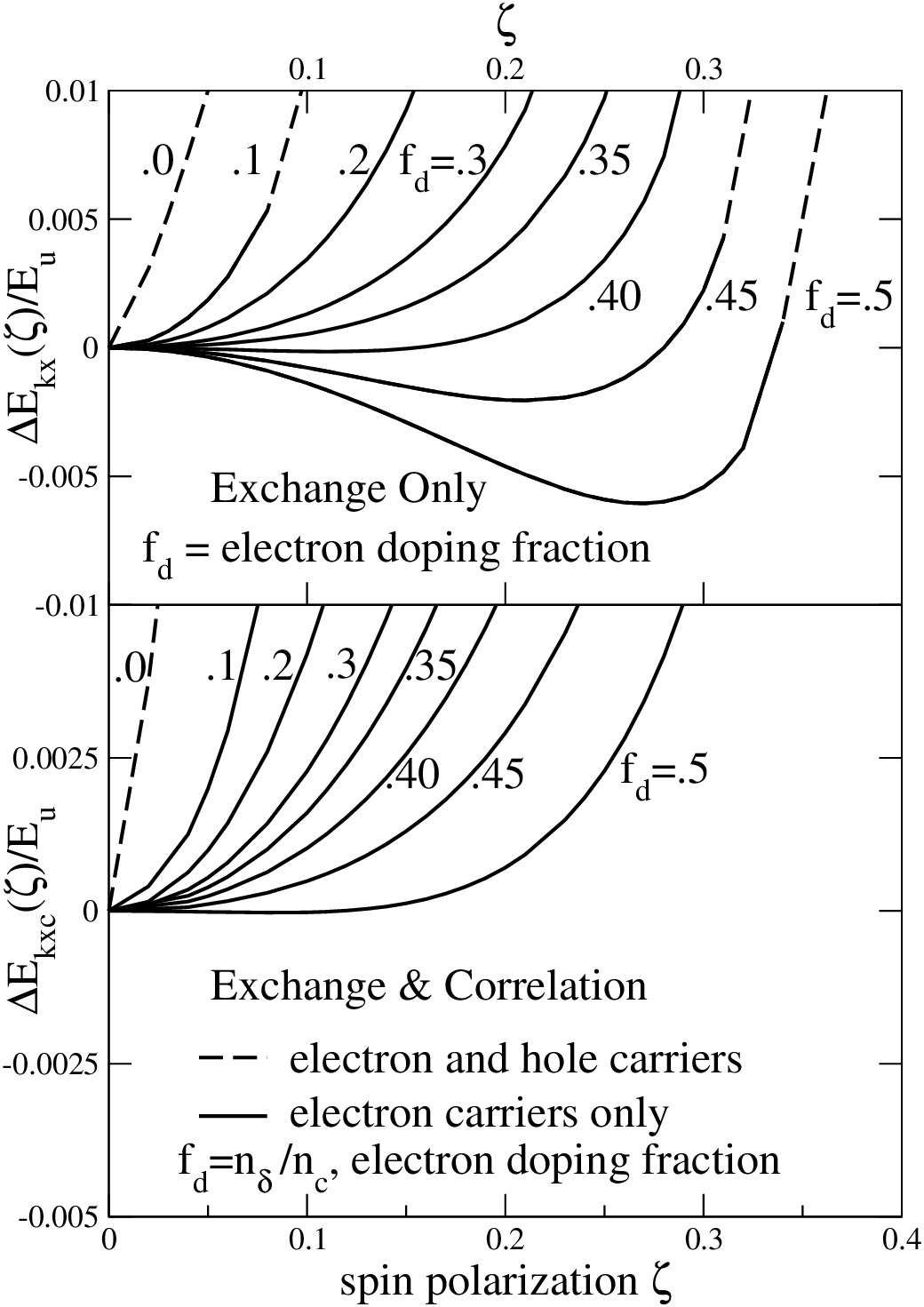}
\caption
{Upper panel-
the energy difference $\Delta E_{kx}$, i.e.,
K.E+exchange,
between the polarized and
unpolarized phases, in units of $E_u=v_Fk_c$, as a function
of the spin polarization $\zeta$ and the dopent fraction
 $f_d$ = $n_\delta/n_c$.
Electron-carrier systems, Fig.~\ref{figcones}(b) are more stable
than electron-hole systems, and show stable spin-polarized states.
However, addition of the correlation energy (lower panel) makes the
unpolarized state the most stable phase. 
}
\label{figexc}
\end{figure}

{\it The kinetic and exchange energies.--}
When the doping per valley is $N_\delta=2n_\delta$, the total number
 of electrons per valley is
$N_t=N_c+N_\delta$. Also, using the $i=1,2,3,4$ notation of
Fig.~\ref{figcones}(b), we set $n_1=n_u$, $n_2=n_d$, $n_3=n_4=n_c$. 
Hence the
spin polarization $s=n_u-b_dn_d$, where the band index $b_d=-1$
for holes. The degree of spin-polarization $\zeta=s/N_t$.
The composition fractions,inclusive of the valley index $v$ =1,2 are
 $x_{vi}=n_i/2N_t$. We note that $k_F=\surd{(2\pi n_\delta)}$,
 $k_u=\surd{\{2\pi(n_\delta+s)\}}$, $k_d=\surd{\{2\pi|n_\delta-s|\}}$.
The exchange energy $E_x(n_\delta,\zeta)$ can be written
as:
\begin{eqnarray}
\label{exceq}
 &&E_{x}(n_\delta,\zeta)/N_t= (N_t/2)\int \frac {2\pi r
dr}{r}\\
\nonumber
&&\times\sum_{ij}x_{vi}x_{vj}[\mathcal G^0_{v,v,ij}(r)-1]
\end{eqnarray}
 It is
implied that the Struve-like component in $\mathcal G_{v,v,ij}$
where $v$ label the valleys, is summed
with the appropriate $b_ib_j$ band $\pm$ factors.
Only a sum over the components in
one valley is needed in evaluating the exchange.
The calculation of both exchange and correlation involves
exactly the same formula, but with a further integration
over a coupling constant $\lambda$ included in the PDFs 
$\mathcal G_{v,v^\prime,ij,\lambda}$. 
The above formula for $E_x$ is made more explicit below.
Thus, the total kinetic and exchange energy $E_{k\,x}$= K.E+$E_x$,
for the case (b) of Fig.~\ref{figcones}
can be written in terms of $n_F, n_u, n_d$
and $n_c$ as in Eq.~\ref{exceq}, or in terms of
 $k_F,k_u,k_d,k_c$ and $A_0$ as:
\begin{eqnarray}
\label{exeq}
 &&E_{kx}(\zeta)=\frac{A_0}{6\pi}v_F(k_u^3+k_d^3)-\\
 \nonumber
&&\frac{A_0}{(2\pi)^2}(g^0v_F/4)(\pi/2)
[k_u^4 \mathcal H_{11}(r)+
\nonumber 
k_d^4 \mathcal H_{22}(r)+\\
\nonumber
&&2k_c^2\{k_u^2\mathcal H_{13}(r)+
k_d^2\mathcal H_{14}(r)\}]
\end{eqnarray}
The kinetic and exchange energy, $E_{t\,x}(\zeta=0)$
of the unpolarized system
shown in Fig.~\ref{figcones}(a), is obtained by
setting $k_F=0$ when  $k_u=k_d$ in Eq.~\ref{exeq}.
 The energy difference
$\Delta E_{k\,x}(\zeta)$ is plotted in the upper panel of
Fig.~\ref{figexc}.
This equation reduces to Eq.(12)-(15) of Ref.~\cite{pgcn} when the
PDFS are replaced by their $k$-space forms and expressed in
terms of elliptic integrals. We have done the calculations
in $r$-space using PDFS, and in $k$-space via
elliptic integrals, as a numerical check.
The evaluations using elliptic intergrals are
numerically more stable. The results (Fig.~\ref{figexc})
show that stable spin-polarized phases appear for systems
containing purely electron carriers,
when the dopant fraction $f_d > 0.38$. 

{\it Effect of the correlation energy $E_c$.--}
This requires the calculation of all the intra and inter-valley
interacting PDFs of G2DES. Even using various symmetries,
this involves calculating nearly two dozen PDFs as well as
an integration over the coupling constant.
We postpone this demanding task to
future work. However, the $E_c(n_\delta,\zeta)$ of G2DES may
be quite well approximated
from the closely analogous four-component 2v-2DES. This
correlation energy/per electron, $\epsilon(r_s,\zeta)$, where
$r_s=g^0$, is given in Eq. 5 of Ref.~\cite{2valley}.
In transferring from the 2v-2DES to the G2DES we note that
$e^2/\epsilon_0$ which is unity in 2v-2DES becomes $g^0v_F$ in
the G2DES. Also, the exchange term arising from the Bessel-like
$h_{11}(r)$, and $h_{22}(r)$,
calculated at the given coupling strength $r_s=g^0$ and
the spin polarization $\zeta$ arise from the same PDFs in
both systems. 
This establishes the scale factor connecting the energy units
of the 2v-2DES and the G2DES at any given spin-polarization.
Given the validity of Eq.~\ref{exceq}, we   adopt the
same scale factor for converting the correlation energy 
$E_c$ of the 2v-2DES to the G2DES.
The energy difference $\Delta E_{k\,xc}(\zeta)$, inclusive of $E_c$, 
between the polarized and
unpolarized phases calculated using the above
approach is shown in the lower panel of Fig.~\ref {figexc}.
The exchange-driven SPP seems to be suppressed to
within the uncertainty of this calculation of $\Delta E_{k\,xc}$.
Improved estimates would only strengthen this calculation further.
 
 In this work we have kept the Coulomb
coupling fixed at $g^0=2.75$ typical of graphene,
 unlike in other studies\cite{tosatti,pgcn} where
the coupling strength $g$ is taken as a tunable parameter,
 (in the spirit of Hubbard-model studies).
 Even in one-valley 2DES, the SPP of low-order theories is
 pushed to $g\sim 26-27$. In the 2v-2DES,
 direct terms (usually known as the correlation terms) predominate over
exchange interactions, and the SPP is {\it suppressed}, as found in
CHNC\cite{2valley} or QMC\cite{conti}.
 Here too, we conclude that the inclusion of correlations
 {\it suppresses the exchange-driven 
spin-phase transition} in the ideal 
 graphene system. %
\end{document}